\documentclass[12pt]{article}
\usepackage{amssymb}

\usepackage[dvips]{graphicx}
\usepackage{epsfig}
\usepackage{amsmath}
\usepackage{float}

\newcommand{\bm}{\begin{multiline}}
\newcommand{\beq}{\begin{equation}}
\newcommand{\eeq}{\end{equation}}
\newcommand{\beqs}{\begin{eqnarray}}
\newcommand{\eeqs}{\end{eqnarray}}

\begin{document}

\thispagestyle{empty}

\begin{flushright}
hep-th/yymmxxx\\
\end{flushright}

\hfill{}

\hfill{}

\hfill{}

\vspace{32pt}

\begin{center}
\textbf{\Large On the gravitational energy of the Kaluza Klein monopole } \\[%
0pt]

\vspace{48pt}

\textbf{Robert B. Mann}\footnote{%
E-mail: rb\texttt{mann@sciborg.uwaterloo.ca}} \textbf{and Cristian Stelea}%
\footnote{%
E-mail: \texttt{cistelea@uwaterloo.ca}}

\vspace*{0.2cm}

\textit{$^{1}$Perimeter Institute for Theoretical Physics}\\[0pt]
\textit{31 Caroline St. N. Waterloo, Ontario N2L 2Y5 , Canada}\\[.5em]

\textit{$^{1,2}$Department of Physics, University of Waterloo}\\[0pt]
\textit{200 University Avenue West, Waterloo, Ontario N2L 3G1, Canada}\\[.5em%
]
\end{center}

\vspace{30pt}

\begin{abstract}
We use local counterterm prescriptions for asymptotically flat space to
compute the action and conserved quantities in five-dimensional Kaluza-Klein
theories. As an application of these prescriptions we compute the mass of the
Kaluza-Klein magnetic monopole. We find consistent results with previous
approaches that employ a background subtraction.
\end{abstract}

\setcounter{footnote}{0}



\section{Introduction}

The problem of defining energy in theories involving gravity has a
long-standing history. One would like for instance to be able to evaluate
the total energy of an isolated object. Throughout the years many
expressions have been proposed for computing the total energy. However,
contrary to initial expectations, it was soon realised that finding
satisfactory quantities is a very difficult task. The essential idea in
computing the energy is to consider the values of the fields far away from
the object and compare them with a background configuration, that is, with a
`no-fields situation'. This is for instance the approach considered when
defining the ADM mass (see for instance \cite{Wald}).

A related problem is that of computing the gravitational action of a
non-compact spacetime. The gravitational action consists of the bulk
Einstein-Hilbert term and it must be supplemented by the boundary
Gibbons-Hawking term in order to have a well-defined variational principle.
When evaluated on non-compact solutions of the field equations it turns out
that both terms diverge. The general remedy for this situation is to
consider the values of these quantities relative to those associated with
some background reference spacetime, whose boundary at infinity has the same
induced metric as that of the original spacetime. The background is chosen
to have a topological structure that is compatible with that of the original
spacetime and also one requires that the spacetimes approaches it
sufficiently rapidly at infinity.

Unfortunately such background subtraction procedures are marred with
difficulties: even if some choices of such reference background spaces
present themselves as `natural', in general these choices are by no means
unique. Moreover, it is not always possible to embed a boundary with a given
induced metric into the reference background and for different boundary
geometries one needs different reference backgrounds \cite{CCM}. A good
example of the difficulties one might encounter in such an endeavour is that
of the celebrated Taub-NUT solution (see for instance \cite{Hawking}-\cite%
{CFM}).

Similar difficulties and ambiguities are encountered when trying to compute
the action and the conserved charges of the Kaluza-Klein monopole \cite%
{sorkin,mp}, and in particular its gravitational energy. Many such
expressions for the conserved charges have been analysed in detail \cite%
{Bombelli,Deser,Onemli:2003gg}; the consistent answers they yield when applied to the
Kaluza-Klein monopole solution are for a definite choice of the reference
background, one that is not a solution of the field equations. Moreover it
is not a flat background, so that the energy expression for a Kaluza-Klein
monopole is problematic. In general potential ambiguities arise in computing
energy and other conserved quantities in dimensionally-reduced gravitation
theories. This is partly because there are many distinct topological
sectors, each of which requires a different background, and partly because
within a given fixed topological sector, there may not be suitable
background.

Motivated by recent results in the AdS/CFT conjecture, Balasubramanian and
Kraus \cite{balakraus} proposed adding a term (referred to as a counterterm)
to the boundary at infinity, which is a functional only of curvature
invariants of the induced metric on the boundary. Such terms will not
interfere with the equations of motion because they are intrinsic invariants
of the boundary metric. By choosing appropriate counterterms, which cancel
the divergences, one can then obtain well-defined expressions for the action
and the energy momentum of the spacetime. Unlike background subtraction,
this procedure is intrinsic to the spacetime of interest and is unambiguous
once the counterterm is specified. While there is a general algorithm for
generating the counterterms for asymptotically (A)dS spacetimes \cite%
{Kraus,GM}, the asymptotically flat case is considerably less-explored (see however \cite{Mann:2005yr} for some new results in this direction).
Early proposals \cite{Lau,Mann1,Ho} engendered study of proposed counterterm
expressions for a class of $(d+1)$-dimensional asymptotically flat solutions
whose boundary topology is $S^{n}\times R^{d-n}$ \cite{Kraus}. This
counterterm method has been applied to the five-dimensional black ring \cite%
{AR} and to an asymptotically Melvin spacetime \cite{Radu}.

The interesting properties of the Kaluza-Klein monopole merit further study
in this context. In the present letter we propose using a local counterterm
prescription to compute its action and its conserved quantities in the
five-dimensional Kaluza-Klein theory. In the next section we introduce the
counterterm action and the expression for the conserved mass using the
boundary stress-energy tensor. In the third section we apply this method to
compute the action and the conserved mass of the Kaluza Klein monopole from
the five-dimensional point of view, while in the fourth section we compute
the monopole energy from the four-dimensional perspective of the
dimensionally reduced theory, using two distinct counterterm prescriptions.
The last section is dedicated to conclusions, in which we comment on the relationships between the various approaches.

\section{The counterterm action}

In $(d+1)$-dimensions, the gravitational action is generally taken to be: 
\begin{equation}
I_{g}=-\frac{1}{16\pi G}\int_{M}d^{d+1}x\sqrt{-g}R-\frac{1}{8\pi G}%
\int_{\partial M}d^{d}x\sqrt{-h}K  \label{actbulkgh}
\end{equation}%
Here $M$ is a $(d+1)$-dimensional manifold with metric $g_{\mu \nu }$, $K$
is the trace of the extrinsic curvature $K_{ij}=\frac{1}{2}h_{i}^{k}\nabla
_{k}n_{j}$ of the boundary $\partial M$ with unit normal $n^{i}$ and induced
metric $h_{ij}$.

For asymptotically flat $4$-dimensional spacetimes, the counterterm 
\begin{equation}
I_{ct}=\frac{1}{8\pi G}\int d^{3}x\sqrt{-h}\sqrt{2\mathcal{R}}
\label{Laumann}
\end{equation}%
was proposed \cite{Lau,Mann1} to eliminate divergences that occur in (\ref%
{actbulkgh}). An analysis of the higher dimensional case \cite{Kraus}
suggested in $5$ dimensions the counterterm 
\begin{equation}
I_{ct}=\frac{1}{8\pi G}\int d^{4}x\sqrt{-h}\frac{\mathcal{R}^{\frac{3}{2}}}{%
\sqrt{\mathcal{R}^{2}-\mathcal{R}_{ij}\mathcal{R}^{ij}}}  \label{I1}
\end{equation}%
where $\mathcal{R}_{ij}$ is the Ricci tensor of the induced metric $h_{ij}$
and $\mathcal{R}$ is the corresponding Ricci scalar. This counterterm
removes divergencies in the action for an asymptotically flat spacetime with
boundary topology $S^{3}\times R$ and also for a $S^{2}\times R^{2}$
boundary topology.

By taking the variation of the action (\ref{I1}) with respect to the
boundary metric $h_{ij}$ we obtain the following boundary stress-energy
tensor: 
\begin{eqnarray}
8\pi G(T_{ct})^{ij}&=&\frac{\mathcal{R}^{\frac{1}{2}}}{\left( \mathcal{R}^2-%
\mathcal{R}_{kl}\mathcal{R}^{kl}\right)^{\frac{3}{2}}}\bigg[ 3\mathcal{R}%
^{ij}\mathcal{R}_{kl}\mathcal{R}^{kl}-\mathcal{R}^{ij}\mathcal{R}^2+2%
\mathcal{R}\mathcal{R}^{ik}\mathcal{R}^j_{~k}+\mathcal{R}^3h^{ij}- \mathcal{R%
}\mathcal{R}_{kl}\mathcal{R}^{kl}h^{ij}\bigg]  \notag \\
&&+\Phi^{(i~;j)k}_{~k}-\frac{1}{2}\Box\Phi^{ij}-\frac{1}{2}%
h^{ij}\Phi^{kl}_{~~;kl},  \notag
\end{eqnarray}
where: 
\begin{eqnarray}
\Phi^{ij}&=&\frac{\mathcal{R}^{\frac{1}{2}}}{\left(\mathcal{R}^2-\mathcal{R}%
_{kl}\mathcal{R}^{kl}\right)^{\frac{3}{2}}}\bigg[2\mathcal{R}\mathcal{R}%
^{ij}+\left(\mathcal{R}^2-3\mathcal{R}_{kl}\mathcal{R}^{kl}\right)h^{ij}%
\bigg],  \notag
\end{eqnarray}
so that the final boundary stress energy tensor is given by: 
\begin{eqnarray}
T_{ij}&=&\frac{1}{8\pi G}\left(K_{ij}-Kh_{ij}+(T_{ct})_{ij}\right)
\label{Tf1}
\end{eqnarray}

For a five-dimensional asymptotically flat solution with a fibred boundary
topology $R^{2}\hookrightarrow S^{2}$, we find that the action (\ref%
{actbulkgh}) can also be regularised using the following equivalent
counterterm 
\begin{equation}
I_{ct}=\frac{1}{8\pi G}\int d^{4}x\sqrt{-h}\sqrt{2\mathcal{R}}  \label{I2}
\end{equation}%
where $\mathcal{R}$ is the Ricci scalar of the induced metric on the
boundary, $h_{ij}$. By taking the variation of this total action with
respect to the boundary metric $h_{ij}$, it is straightforward to compute
the boundary stress-tensor, including (\ref{I2}): 
\begin{equation*}
T_{ij}=\frac{1}{8\pi G}\left( K_{ij}-Kh_{ij}-\Psi( \mathcal{R}_{ij}-\mathcal{%
R}h_{ij})-h_{ij}\Box\Psi+\Psi_{;ij}\right)
\end{equation*}
where we denote $\Psi=\sqrt{\frac{2}{\mathcal{R}}}$. If the boundary
geometry has an isometry generated by a Killing vector $\xi ^{i}$, then $%
T_{ij}\xi ^{j}$ is divergence free, from which it follows that the quantity 
\begin{equation*}
\mathcal{Q}=\oint_{\Sigma }d^{3}S^{i}T_{ij}\xi ^{j},
\end{equation*}%
associated with a closed surface $\Sigma $, is conserved. Physically, this
means that a collection of observers on the boundary with the induced metric 
$h_{ij}$ measure the same value of $\mathcal{Q}$, provided the boundary has
an isometry generated by $\xi $. In particular, if $\xi ^{i}=\partial
/\partial t$ then $\mathcal{Q}$ is the conserved mass $\mathcal{M}$.

The counterterm (\ref{I1}) was proposed in \cite{Kraus} for spacetimes with
boundary $S^{2}\times R^{2}$, or $S^{3}\times R$. On the other hand, the
counterterm (\ref{I1}) is essentially equivalent to (\ref{I2}) for $%
S^{2}\times R^{2}$ boundaries. We find that when the boundary is taken to
infinity both expressions cancel the divergences in the action. Our choice
of using (\ref{I2}) can be motivated by the fact that the expression for the
boundary stress-tensor is considerably simpler. However, different
counterterms can lead to different results when computing the energy,
seriously constraining the various choices of the boundary counterterms (see
for instance \cite{Hollands1,Hollands2} for a general study of the
counterterm charges and a comparison with charges computed by other means in 
$AdS$ context). As we shall see in the next section, both expressions lead
to a background-independent Kaluza-Klein mass that agrees with other answers
previously known in the literature; however, we do find slight discrepancies
in the diagonal components of the boundary stress-tensor .

\section{The mass of the Kaluza-Klein Monopole}

We begin by reviewing the original magnetic monopole solution in four
dimensions that arises as a Kaluza-Klein compactification of a five
dimensional vacuum metric \cite{sorkin,mp} (see also \cite{Mann2}). The
essential ingredient used in the monopole construction is a four-dimensional
version of the Taub-NUT solution, with Euclidean signature. To construct the
monopole solution, we start with the Euclidean form of the Taub-NUT solution %
\cite{Taub,NUT,Misner}: 
\begin{equation*}
ds^{2}=F_{E}(r)(d\chi -2n\cos \theta d\varphi
)^{2}+F_{E}^{-1}(r)dr^{2}+(r^{2}-n^{2})d\Omega ^{2}
\end{equation*}%
where 
\begin{equation}
F_{E}(r)=\frac{r^{2}-2mr+n^{2}}{r^{2}-n^{2}}  \label{TN}
\end{equation}%
In general, the $U(1)$ isometry generated by the Killing vector $\frac{%
\partial }{\partial \chi }$ (that corresponds to the coordinate $\chi $ that
parameterizes the fibre $S^{1}$) can have a zero-dimensional fixed point set
(referred to as a `nut' solution) or a two-dimensional fixed point set
(correspondingly referred to as a `bolt' solution). The regularity of the
Euclidean Taub-nut solution requires that the period of $\chi $ be $\beta
=8\pi n$ (to ensure removal of the Dirac-Misner string singularity), $%
F_{E}(r=n)=0$ (to ensure that the fixed point of the Killing vector $\frac{%
\partial }{\partial \chi }$ is zero-dimensional) and also $\beta
F_{E}^{\prime }(r=n)=4\pi $ in order to avoid the presence of the conical
singularities at $r=n$. With these conditions we obtain $m=n$, yielding 
\begin{equation}
F_{E}(r)=\frac{r-n}{r+n}  \label{FEnut}
\end{equation}%
Taking now the product of this Euclidean space-time with the real line, we
obtain the Kaluza-Klein monopole, as described by the following
five-dimensional Ricci flat metric: 
\begin{equation}
ds^{2}=-dt^{2}+F_{E}(r)(d\chi -2n\cos \theta d\varphi
)^{2}+F_{E}^{-1}(r)dr^{2}+(r^{2}-n^{2})d\Omega ^{2}
\end{equation}%
The other possibility to explore is using the Taub-bolt solution in
four-dimensions instead of the nut solution. In this case the Killing vector 
$\frac{\partial }{\partial \chi }$ has a two-dimensional fixed point set in
the four-dimensional Euclidean sector. The regularity of the solution is
then ensured by demanding that $r\geq 2n$, while the period of the
coordinate $\chi $ is $8\pi n$ and for the bolt solution we obtain (with $%
m=5n/2$) \cite{Page}: 
\begin{equation}
F_{E}(r)=\frac{\left( r-2n\right) \left( r-\frac{1}{2}n\right) }{r^{2}-n^{2}}
\label{FEbolt}
\end{equation}%
As in the case of the nut solution, we take the product with the real line
and obtain a metric in five-dimensions that is a solution of the vacuum
Einstein field equations. The physical interpretation of this last solution
was recently clarified by Liang and Teo \cite{Edward1}. It corresponds to a
pair of coincident extremal dilatonic black holes with opposite unequal
magnetic charges.

Before we apply the counterterm prescription to compute the components of
the boundary stress-tensor, let us notice that the boundary topology of the
KK monopole for constant, finite values of the radial coordinate $r$ is that
of a squashed $3$-sphere times a real line. Therefore, one might expect that
the proper counterterm action to use should be the one corresponding to an $%
S^{3}\times R$ topology. However using that counterterm we find that it is
impossible to cancel out the divergences as $r\rightarrow \infty $. Rather
we note that, as $r\rightarrow \infty $, the boundary topology is that of a
fibre bundle $R\times S^{1}\hookrightarrow S^{2}$ as the radius of $S^{2}$
grows with $r$, while the radius of $S^{1}$ reaches a constant value.
Thence, asymptotically, the choice of the counterterm (\ref{I2}) is natural
and indeed, we find that using this counterterm we can eliminate the
divergences in the action and obtain finite values for the total mass.

Using the metric with the general expression (\ref{TN}) for the function $%
F_{E}(r)$ we find 
\begin{eqnarray}
8\pi GT_{~t}^{t} &=&\frac{m}{r^{2}}+O(r^{-3})  \notag \\
8\pi GT_{~\chi }^{\chi } &=&\frac{2m}{r^{2}}+O(r^{-3})  \notag \\
8\pi GT_{~\phi }^{\chi } &=&\frac{4mn\cos \theta }{r^{2}}+O(r^{-3})  \notag
\\
8\pi GT_{\theta }^{\theta } &=&-\frac{m^{2}-2n^{2}}{2r^{3}}+O(r^{-4})  \notag
\\
8\pi GT_{\phi }^{\phi } &=&-\frac{m^{2}-2n^{2}}{2r^{3}}+O(r^{-4})
\label{TKK2}
\end{eqnarray}%
the rest of the terms being of order $O(r^{-3})$ or higher. Then the
conserved mass associated with the Killing vector $\xi =\partial /\partial t$
is found to be: 
\begin{equation*}
\mathcal{M}=\frac{4\pi mn}{G}
\end{equation*}%
However using the counterterm (\ref{I1}) and the boundary stress-tensor (\ref%
{Tf1}) we obtain%
\begin{eqnarray}
8\pi GT_{~t}^{t} &=&\frac{m}{r^{2}}+O(r^{-3})  \notag \\
8\pi GT_{~\chi }^{\chi } &=&\frac{2m}{r^{2}}+O(r^{-3})  \notag \\
8\pi GT_{~\phi }^{\chi } &=&\frac{4mn\cos \theta }{r^{2}}+O(r^{-3})  \notag
\\
8\pi GT_{\theta }^{\theta } &=&-\frac{m^{2}-4n^{2}}{2r^{3}}+O(r^{-4})  \notag
\\
8\pi GT_{\phi }^{\phi } &=&-\frac{m^{2}-4n^{2}}{2r^{3}}+O(r^{-4})
\label{TKK1}
\end{eqnarray}%
It is easy to see that this boundary stress-energy tensor leads to the same
mass as above. Notice however that some of the components of the
stress-energy tensor (\ref{TKK1}) are different from the ones obtained in (%
\ref{TKK2}).

For Kaluza-Klein monopole we have $m=n$ and we obtain $\mathcal{M}=\frac{%
4\pi n^{2}}{G}$, which is easily seen to be the same with the one derived in %
\cite{Bombelli,Deser} by using a background subtraction procedure.\footnote{%
The parameter $\lambda _{\infty }$ used in \cite{Bombelli} corresponds in
our case to $4n$, while $k=8\pi G$.} For the bolt monopole we have $m=5n/2$
and using either prescription (\ref{I1})\ or (\ref{I2}) we obtain $\mathcal{M%
}=\frac{10\pi n^{2}}{G}$. In both cases the regularized action takes the
form $I=\beta \mathcal{M}$, where $\beta $ is the periodicity of the
Euclidian time $\tau =it$. Upon application of the Gibbs-Duhem relation $%
S=\beta \mathcal{M}-I$ we find that the entropy is zero, as expected since
there are no horizons.

\section{The monopole mass from the four dimensional perspective}

It is instructive to compute the conserved mass after we perform the
dimensional reduction along the $\chi $ direction down to four-dimensions.
While both the metric and the fields in general have singularities at the
origin, this is not necessarily an obstruction since the conserved charges
are in general computed as surface integrals at infinity.

Using the metric ansatz: 
\begin{equation*}
ds_{5d}^{2}=e^{\frac{\varphi }{\sqrt{3}}}ds_{4d}^{2}+e^{\frac{-2\varphi }{%
\sqrt{3}}}(d\chi +\mathcal{A})^{2}
\end{equation*}%
we obtain the four-dimensional fields: 
\begin{eqnarray}
ds_{4d}^{2} &=&-F_{E}^{\frac{1}{2}}dt^{2}+F_{E}^{-\frac{1}{2}%
}(r)dr^{2}+F_{E}^{\frac{1}{2}}(r^{2}-n^{2})d\Omega ^{2}  \notag \\
\mathcal{A} &=&-2n\cos \theta d\varphi ,~~~~~~~e^{\frac{\phi }{\sqrt{3}}%
}=F_{E}^{-\frac{1}{2}}  \label{sorkinmonopole}
\end{eqnarray}%
It is clear that the metric is asymptotically flat and the form of the
electromagnetic potential $\mathcal{A}$ describes the electromagnetic field
generated by a magnetic monopole.

In four-dimensions we can use the counterterm (\ref{Laumann}) , whose only
difference from (\ref{I2})\thinspace is that we are integrating now over a
three-dimensional boundary instead of a four-dimensional one. A similar
computation with the one performed in five-dimensions yields 
\begin{eqnarray}
8\pi G_{4}T_{t}^{t} &=&\frac{m}{r^{2}}+O(r^{-3})  \notag \\
8\pi G_{4}T_{\theta }^{\theta } &=&\frac{n^{2}-m^{2}}{2r^{3}}+O(r^{-4}) 
\notag \\
8\pi G_{4}T_{\phi }^{\phi } &=&\frac{n^{2}-m^{2}}{2r^{3}}+O(r^{-4})
\end{eqnarray}%
for boundary stress-energy tensor, where $G_{4}$ is Newton's constant in
four-dimensions. Then the conserved mass associated with the Killing vector $%
\xi =\partial /\partial t$ is found to be: 
\begin{equation*}
\mathcal{M}=\frac{m}{2G_{4}}
\end{equation*}%
Noting that we have the relation $G=8\pi nG_{4}$ we find that the mass
computed using the four-dimensional geometry agrees precisely with the one
computed in the five-dimensional theory.

Finally, we shall compute the mass using the methods from \cite{Mann:2005yr}. In that work, Mann and Marolf put forward a new counterterm that is also
given by a local function of the boundary metric and its curvature tensor.
The new counterterm is taken to be the trace $\hat{K}$\ of a symmetric
tensor $\hat{K}_{ij}$ that is defined implicitly in terms of the Ricci
tensor $\mathcal{R}_{ij}$ of the induced metric on the boundary via the
relation 
\begin{equation}
\mathcal{R}_{ik}=\hat{K}_{ik}\hat{K}-\hat{K}_{i}^{m}\hat{K}_{mk},
\label{MannMarolf}
\end{equation}%
In contrast to previous counterterm proposals (such as (\ref{I1})) this new
counterterm assigns an identically zero action to the flat background in any
coordinate systems while giving finite values for asymptotically flat
backgrounds. The renormalized action leads to the usual conserved quantities
that can also be expressed in terms of a boundary stress-tensor whose
expression involves the electric part of the Weyl tensor:\footnote{%
Even if the four-dimensional solution is not a vacuum metric, the net effect
of the matter fields is to give only sub-leading order corrections and to
leading order we can still replace the bulk Riemann tensor with the Weyl
tensor.} 
\begin{equation*}
T_{ij}^{0}u^{j}=\frac{1}{8\pi G_{4}}rE_{ij}u^{j}
\end{equation*}%
Here $E_{ij}$ is the pull-back to the boundary of the contraction of the
bulk Weyl tensor with the induced metric while $u^{i}$ is the normal to the
spacelike surface $\Sigma $. Computing this expression in the $r\rightarrow
\infty $ limit and contracting with the Killing vector $\xi =\partial
/\partial t$ we obtain: 
\begin{equation*}
T_{ij}^{0}\xi ^{i}u^{j}=\frac{1}{8\pi G_{4}}\frac{m}{r^{2}}+O(r^{-3})
\end{equation*}%
while the conserved mass is found by simple integration to be: 
\begin{equation*}
\mathcal{M}=\frac{m}{2G_{4}}
\end{equation*}%
in agreement with previous computations.

\section{Discussion}

In General Relativity there are many known expressions for computing the
energy in asymptotically flat spacetimes. The general idea is to study the
asymptotic values of the gravitational field, far away from an isolated
object, and compare them with those corresponding to a gravitational field
in the absence of the respective object. However, most of these proposals
will provide results that are relative to the choice of a reference
background (be it a spacetime metric or merely a connection). The background
must be chosen such that its topological properties match the solution whose
action and conserved charges we want to compute. However, this does not fix
the choice of the background and moreover, there might be cases in which the
topological properties of the solution rule out any natural choice of the
background.

Most of these difficulties are simply avoided once we resort to the
counterterm-method \cite{balakraus,Kraus,GM,Lau,Mann1}. The main
motivation for the present work was to investigate the local counterterm
prescription for computing the action and the conserved charges in the
five-dimensional Kaluza-Klein theory  and, more specifically, for the
Kaluza-Klein monopole solution. The main advantage of this approach is that
it gives results that are intrinsic to the solutions considered, that is,
the results are not `relative' to some reference background. Using
two distinct proposals for the boundary counterterm we computed the mass of
the Kaluza-Klein magnetic monopole and found agreement in both cases with
previous results derived by other means \cite{Bombelli,Deser}. We also
extended our results to the case of the Kaluza-Klein bolt-monopole solution.
In the general context of Kaluza-Klein theory it is also tempting to examine
the energy from the point of view of the dimensionally-reduced theory. While
the metric and also the fields do have in general singularities at origin,
this is not necessarily an obstruction since the conserved charges are in
general computed as surface integrals at infinity. In the four-dimensional
theory, using the counterterm (\ref{Laumann}) proposed by Lau \cite{Lau} and
Mann \cite{Mann1} as well as the new counterterm proposed in \cite{Mann:2005yr} we computed the mass of the monopole and found it to be
equal to the five-dimensional mass. A similar result was proved in \cite%
{Bombelli} using background subtraction methods.

Let us remark that the counterterm method for computing conserved charges
might shed some light on the old problem of which compactifications are
preferred in Kaluza-Klein theories. This problem involves a comparison of
the gravitational energies corresponding to different vacua. The advantage
of the counterterm method is that by providing results that are intrinsic to
spacetime geometries it obviates the need to consider only the solutions
corresponding to the same asymptotic reference background.

Finally, we believe that our results warrant further study of the
counterterm method in asymptotically flat spacetimes. That the conserved charges computed from these various counterterm-supplemented actions 
agree is not surprising in view of the results of refs. \cite{Mann:2005yr,Hollands1,Hollands2}. 
The actions associated with the three counterterm prescriptions (\ref{Laumann},\ref{I1},\ref{MannMarolf})
all lead to well-defined variational principles (as shown on general grounds in \cite{Mann:2005yr}) and the actions are all finite on all solutions with the asymptotics of the Kaluza-Klein monopole. 
Consequently the energies computed
from the various approaches can only differ by c-number terms \cite{Hollands2}. However, while it is clear that the distinct choices (\ref{I1}) and (\ref{I2}) yield the same mass for the
monopole, the diagonal components of the boundary stress-energy tensor have
slightly different coefficients. The implications of this remain an
interesting subject for future investigation.
\medskip

{\Large Acknowledgements}

This work was supported by the Natural Sciences and Engineering Council of
Canada.

\end{document}